\documentclass[aps,pra,floatfix,twocolumn,superscriptaddress,showpacs,10pt]{revtex4-1}
\usepackage{amssymb,amsmath,color,mciteplus,graphicx,subfigure}
\usepackage{calc,epsfig,epstopdf,color,mciteplus,bm,mathrsfs}
\usepackage[utf8]{inputenc}
\usepackage{times}
\usepackage{appendix}

\usepackage{lipsum}
\usepackage[normalem]{ulem}
\usepackage{xcolor}
\newcommand{\del}[1]{}
\newcommand{\add}[1]{#1}

\usepackage{color}
\usepackage{amsmath}
\usepackage{braket}
\usepackage[unicode=true,
bookmarks=true,bookmarksnumbered=false,bookmarksopen=false,
breaklinks=false,pdfborder={0 0 1},backref=false,colorlinks=true]
{hyperref}
\hypersetup{linkcolor=magenta, urlcolor=blue, citecolor=blue,pdfstartview={FitH}, hyperfootnotes=false, unicode=true}

\usepackage{float}
\begin{document}

\title{Multi-flux Aharonov-Bohm caging with tunable couplings}

\author{Le-Chuan Wang}
\affiliation{Key Laboratory of Atomic and Subatomic Structure and Quantum Control (Ministry of Education), Guangdong Basic Research Center of Excellence for Structure and Fundamental Interactions of Matter, and School of Physics, South China Normal University, Guangzhou 510006, China}

\author{Sai Li}

\affiliation{Key Laboratory of Atomic and Subatomic Structure and Quantum Control (Ministry of Education), Guangdong Basic Research Center of Excellence for Structure and Fundamental Interactions of Matter, and School of Physics, South China Normal University, Guangzhou 510006, China}

\affiliation{Guangdong Provincial Key Laboratory of Quantum Engineering and Quantum Materials, Guangdong-Hong Kong Joint Laboratory of Quantum Matter, and  Frontier Research Institute for Physics,\\ South China Normal University, Guangzhou 510006, China}

\author{Jia Liu}\email{liujia@quantumsc.cn}

\affiliation{Quantum Science Center of Guangdong-Hong Kong-Macao Greater Bay Area, Shenzhen 518045, China}

\author{Zheng-Yuan Xue}\email{zyxue83@163.com}
\affiliation{Key Laboratory of Atomic and Subatomic Structure and Quantum Control (Ministry of Education), Guangdong Basic Research Center of Excellence for Structure and Fundamental Interactions of Matter, and School of Physics, South China Normal University, Guangzhou 510006, China}

\affiliation{Guangdong Provincial Key Laboratory of Quantum Engineering and Quantum Materials, Guangdong-Hong Kong Joint Laboratory of Quantum Matter, and  Frontier Research Institute for Physics,\\ South China Normal University, Guangzhou 510006, China}

\affiliation{Quantum Science Center of Guangdong-Hong Kong-Macao Greater Bay Area, Shenzhen 518045, China}

\affiliation{Hefei National Laboratory, Hefei 230088, China}
\date{\today}

\begin{abstract}
Aharonov-Bohm (AB) caging is the complete wavefunction localization effect in translational-invariant lattices induced by destructive phase interference. These phases originate from the gauge fields such as the penetrated magnetic fields, which are directly related to several novel topological quantum states of matter. Recently, this effect has shown significant potential for applications in quantum simulation and topological quantum computation. Here, we propose a scalable protocol to derive universal conditions for AB caging with multi-flux. The numerical simulations validate the theoretical predictions by directly observing AB caging phenomena. We also investigate the breakage of the caging effect with onsite detuning, decoherence and non-Hermitian transitions. \del{Our protocol can be directly tested in several quantum many-body platforms and provides an alternative approach for advancing quantum simulation of exotic state matter.} \add{Our protocol provides an alternative approach for advancing quantum simulation of exotic states of matter.}
\end{abstract}

\maketitle

\section{Introduction}

Since P. W. Anderson discovered localization phenomena  in lattices \cite{coupling2}, research on localization in many-body systems has remained a subject of sustained interest. With advances in experimental capabilities, theoretical and experimental studies on the Anderson localization in various physical platforms have garnered increasing attention in recent years \cite{AW0,AW1,AW2,AW3}. The Anderson localization is achieved through doping-induced disorder, however there are also other mechanisms that can cause localized phenomena. One of the well-known examples is the Aharonov-Bohm (AB) caging effect, which is characterized by the complete confinement of  particles' wavefunction within a specific periodic lattice. And the wave-function's transport  is significant suppressed \cite{transport1, transport2, transport3}. This localization phenomenon is triggered by quantum interference effects induced by gauge fields \cite{Caging1, Caging2, Caging3}. It has recently attracted considerable attention due to its potential applications in quantum simulation and topological quantum computation \cite{topological_quantum_computation}. For example, in photonic lattices, a similar mechanism has been utilized to design dissipationless light transport pathways \cite{photonic_lattice1}. On superconducting circuits \cite{superconducting_circuits}, the realization of AB caging offers a new platform for investigating correlated topological matter.

However, there are still many unexplored aspects. For photonic lattice systems \cite{transport2,photonic_lattice1,photonic_lattice2}, the inevitable optical loss and dephasing effect in the waveguide array may affect long-distance transmission observation. Only the AB caging effect with two fluxes is experimentally demonstrated in superconducting circuit systems. The decoherence time of superconducting qubits limits the evolution observation, and the complexity of the hardware is a great challenge in experiment. For the ultra-cold atom systems \cite{Ultracold_atom1, Ultracold_atom2, Ultracold_atom3, Ultracold_atom4, Ultracold_atom5, Ultracold_atom6, Ultracold_atom7, Ultracold_atom8}, the complex Raman-coupled lattice in the experiment affects the observation of quantum transport, which significantly slows the dynamical process especially under the strong disorder condition. On the other hand, the small asymmetry in the control of coupling strength and gauge field, leads
to leakage in the AB caging \cite{leakage}. The trapped-ion system is a quantum model, where the Coulomb interaction between ions exists naturally, allows one to explore the coupling effect of AB cage with long-range correlation \cite{trapped_ion1}. In addition, the longer quantum state coherence time in the ion trap system makes it more favorable to observe the long-time evolution dynamics and the stability of the high-dimensional state compared to other systems \cite{trapped_ion2, trapped_ion3}.

Here, we propose a protocol to derive universal AB caging conditions in multi-flux lattices with tunable coupling. This work first shows the possible feasibility of the proposed method. It may be suitable for future research on multi-level trapped ion systems \cite{linyh} and superconducting circuits \cite{xue} with parametric tunable couplings. Starting from a one-dimensional (1D) translational-invariant lattice with multi-flux, we first establish flux-localization relationships and give a special example with numerical verifications. We further investigate the robustness of the caging effect by introducing on-site potentials and decoherence effect. At last we explore the influence of non-Hermitian transitions on AB caging.

\begin{figure}[tbp]
    \includegraphics[width=1\linewidth]{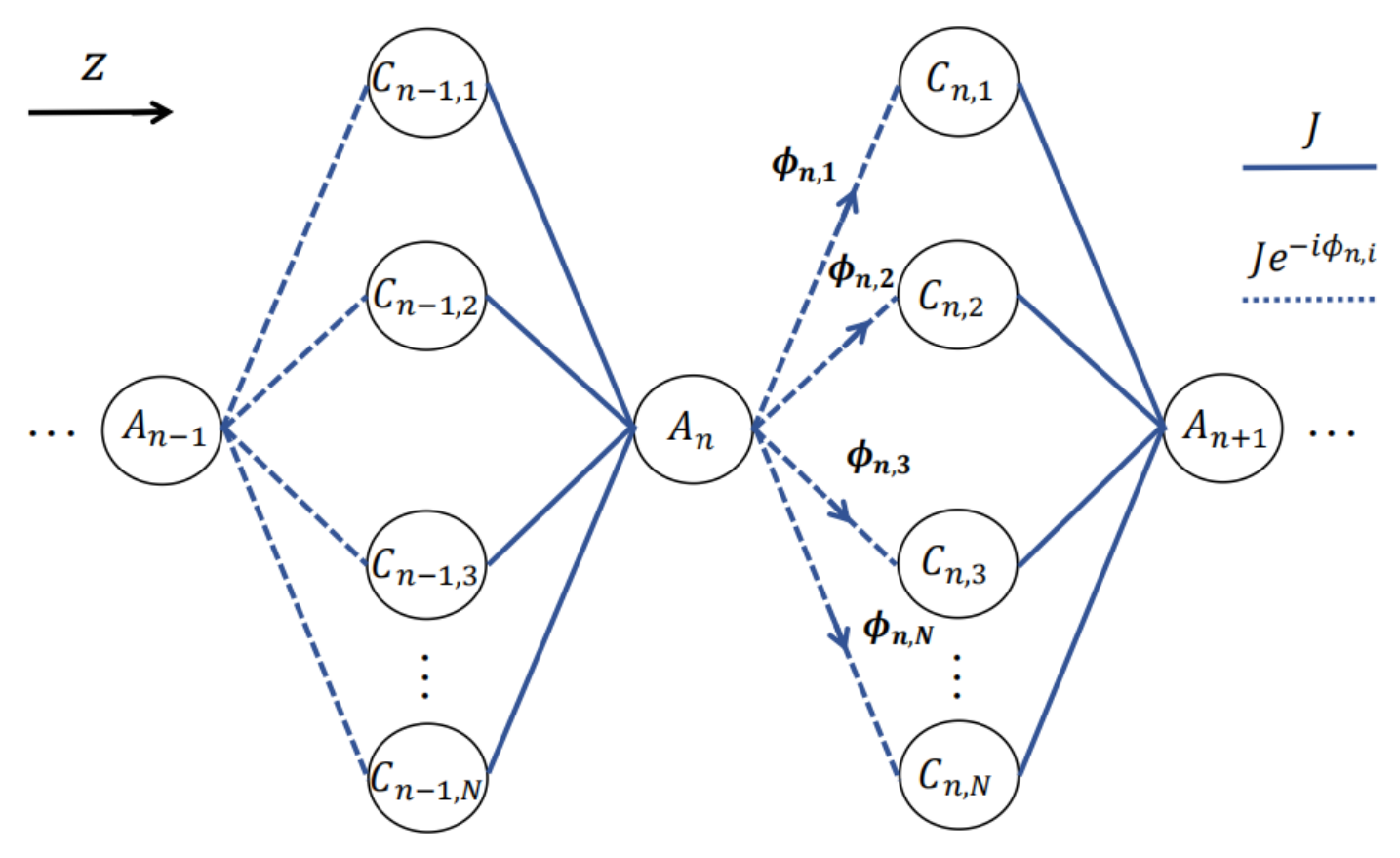}
        \caption{The interaction between the lattice sites $A_{n}$, $C_{n,i}$ and $A_{n+1}$ is shown in the figure, where the solid line represents the strength to be a constant $J$. The dashed line indicates that there is a transition phase $e^{-i\phi_{n,i}}$ in the coupling between the lattice, and there are $N$ paths($C_{n,i}$ ) between the adjacent sites $A_{n}$ and $A_{n+1}$. The transmission process of photons in the lattice is along the $z$ direction.}
    \label{f1}
\end{figure}

\section{multi-flux lattice model}
 Consider a one-dimensional multi-flux lattice system, as shown in Fig. \ref{f1}, the unit cells of which are composed of the lattice sites $A_{n}$ and $C_{n,i}$. The Hamiltonian of this system can be written as
\begin{equation}     \label{Hsum}
    \hat{H}=\sum_{n}\sum_{i=1}^{N}\left ( Je^{-i\phi _{i}}\hat{A}_{n}^{\dagger}\hat{C}_{n,i}+J\hat{A}_{n}^{\dagger}\hat{C}_{n-1,i}+\text{H.c.} \right ),
\end{equation}
where $J$ is the coupling strength, $\phi_{i}$ is the transition phase and $\hat{A}_{n}$ and $\hat{C}_{n,i}$ are annihilation operators. For simplicity, we assume that the coupling strength between the lattice sites is all constant and has a magnitude of $J$. Transffering to the momentum($\textit{k}$) space we can obtain
\begin{equation} \label{H_k}
    \hat{\mathcal{H}}\left ( k,\phi_{i} \right )=\sum_{k}\sum_{i=1}^{N}J_{i}\hat{A}_{k}^{\dagger}\hat{C}_{i,k}+\text{H.c.},
\end{equation}
where \(\hat{A}_{k}=\sum_{n}\hat{A}_{n}e^{-ikn}\), \(\hat{C}_{k,i}=\sum_{n}\hat{C}_{n,i}e^{-ikn}\), the coupling strength in \textit{k} space satisfies \( J_{i}=J\left ( e^{-i\phi_{i}}+e^{-ik} \right ) \), reflects the modulation of the phase $\phi_{i}$ and the momentum \textit{k} to the coupling. Accordingly, the Hamiltonian of this system in \textit{k} space in matrix form is,
\begin{equation} \label{Hsum matrix}
    \mathcal{H}=\begin{pmatrix}
     0&J_{1}&J_{2}&.&.&.&J_{N}\\
    J^{*}_{1}&0&0&.&.&.&0\\
    J^{*}_{2}&0&0&.&.&.&0\\
    .&.&.&.&&&.\\
     .&.&.&&.&&.\\
      .&.&.&&&.&.\\
      J^{*}_{N}&0&0&.&.&.&0
\end{pmatrix}.
\end{equation}
By solving the eigenvalue equation we can find two eigenstates of the above Hamiltonian as
\begin{equation}\label{eigenstates}
    \left| \psi_{\pm}\left ( k \right ) \right\rangle=\frac{1}{\sqrt{2}}\left [ \left| A_{k} \right\rangle\pm\frac{\sum_{i}J_{i}\left| C_{i,k} \right\rangle}{E_{k}} \right ],
\end{equation}
where $\left| A_{k} \right\rangle$ and $\left| C_{i,k} \right\rangle$ are Bloch mode basis, and eigenenergies $E_{k}$ can be written as
 \begin{equation}\label{E_{k}}
    E_{k}=\sqrt{\sum_{i=1}^{N}2J^{2}\left ( 1+\mathrm{cos}\phi_{i} \mathrm{cos}k+\mathrm{sin}\phi_{i} \mathrm{sin}k \right )}\quad.
 \end{equation}

 The localization of particles in the lattice can be understood using the group velocity $v_{g}\propto \mathrm{d}E_{k}/\mathrm{d}k$ to describe the propagation behavior of wave functions. When the energy eigenvalues satisfy $\mathrm{d}E_{k}/\mathrm{d}k\neq 0$, the energy is dispersive, and the group velocity $v_{g}$ is not zero. That represents the normal particle transport in the lattice. When $\mathrm{d}E_{k}/\mathrm{d}k=0$ for any $k$, the lattice is flat-band with no energy dispersive. The group velocity is zero and there is no particle transport in the lattice\cite{Vg}. In conventional flat-band systems, localized states originate from the geometry and connectivity of the lattice and can not be freely tunable \cite{flat_band}. In contrast, for AB caging, localized states originate from destructive interference among different paths. Through artificial gauge fields, the flux $\phi_{n,i}$ and path number $N$ can be flexibly modulated to achieve localized states. Moreover, gradual tuning of $\phi_{n,i}$ within a certain range enables observation of the transition between localized and extended states in the system. AB caging arises from a flat band with $E_k$ as a constant for any \textit{k}, which can be achieved under the conditions of
\begin{equation} \label{condition}
\sum_{i=1}^{N}\mathrm{cos}\phi_{i}=0,\quad
\sum_{i=1}^{N}\mathrm{sin}\phi_{i}=0.
\end{equation}

For the phenomenon of localization, the two equations in Eq. (\ref{condition}) need to be established at the same time. Considering the transport process of photons in lattice \cite{transport_process1}, there are $N$ paths for the transport from one lattice point to another. For the $N$ paths case, two special examples are given below. When $N$ is odd, symmetric phase configurations satisfy
\begin{equation}\label{odd}
\begin{aligned}
\phi_{1}&=0,\ \phi_{2}=\phi_{3}=...=\phi_{\frac{N-1}{2}}=\phi,\\
\phi_{\frac{N+1}{2}}&=\phi_{\frac{N+3}{2}}=\phi_{\frac{N+5}{2}}=...=\phi_{N}=-\phi,
\end{aligned}
\end{equation}
substituting Eq. (\ref{odd}) into Eq. (\ref{condition}), we can find that the transition flux corresponding to the occurrence of the cage effect satisfies
\begin{equation}\label{odd condition}
\phi=\mathrm{arccos}\left ( -\frac{1}{N-1} \right ).
\end{equation}
When $N$ is even, symmetric phase configurations satisfy
\begin{equation}\label{even phi}
\begin{aligned}
 \phi_{1}&=\phi_{2}=...=\phi_{\frac{N}{2}}=\phi,\\
 \phi_{\frac{N}{2}+1}&=\phi_{\frac{N}{2}+2}=...=\phi_{N}={\phi}'.
\end{aligned}
\end{equation}
By the same token, we can see that the transition fluxes which suffer from the caging effect satisfy
\begin{equation}\label{even condition}
{\phi_{i}}'=\phi_{i}+m\pi\left ( m=1,3,5... \right ).
\end{equation}
For simplicity, we set $m = 1$ to obtain
\begin{equation}\label{even condition}
    \begin{aligned}
 \phi_{1}&=\phi_{2}=...=\phi_{\frac{N}{2}}=\phi,\\
 \phi_{\frac{N}{2}+1}&=\phi_{\frac{N}{2}+2}=...=\phi_{N}=\phi+\pi.
\end{aligned}
\end{equation}
\section{  Localization Analysis via Inverse Participation Number}

Now we consider the light dynamics in the different caging regimes. As discussed by others, assuming evanescent coupling of single-mode waveguides \cite{coupling1, coupling2}, it is described by the following coupled-mode equations. In our system, The dynamics of particles in the lattice are determined by the Schr\"odinger equation
 \begin{equation}\label{coupledmode equations}
\begin{cases}
i\partial_{t}a_{n} &=J_{i}\sum_{i=1}^{N} e^{-i\phi_{i}}c_{n,i}+J_{i}\sum_{i=1}^{N} c_{n-1,i},\\
          i\partial_{t}c_{n,1} &=J_{i}e^{i\phi_{1}}a_{n}+J_{i}a_{n+1},\\
         i\partial_{t}c_{n,2} &=J_{i}e^{i\phi_{2}}a_{n}+J_{i}a_{n+1},\\
        &\;.\;.\;.\\
         &\;.\;.\;.\\
         i\partial_{t}c_{n,N} &=J_{i}e^{i\phi_{N}}a_{n}+J_{i}a_{n+1},
\end{cases}
\end{equation}
where $a_{n}$ and $c_{n,i}$ represent the probability amplitude of photons appearing at each lattice  in the system. To better characterize the photon localization properties in the lattice, we define the inverse participation number (IPN), which serves as a fundamental metric for quantifying wavefunction localization in disordered or periodic systems. It is
 \add{\begin{equation}\label{p^{-1}}
p^{-1}\left ( t \right )=\sum_{j=1}^{m}\left | \psi_{j}\left ( t \right ) \right |^{4},
\end{equation}}
\del{where $\psi_{i}\left ( t \right )$ represents the wavefunction amplitude at mode $i$ and $N$ is the total number of modes.} \add{where $\psi_{j}\left ( t \right )$ represents the wavefunction amplitude at mode $j$ and $m$ is the total number of modes.} The IPN's physical significance emerges from its bounding properties, under the constraint of normalization conditions \del{$\sum_{i=1}^{N}\left | \psi_{i}\left ( t \right ) \right |^{2}=1$} \add{$\sum_{j=1}^{m}\left | \psi_{j}\left ( t \right ) \right |^{2}=1$}, combining the Schwartz inequality gives an upper bound on the inverse participation number of
\add{\begin{equation}\label{p^{-1}upper bound}
\sum_{j=1}^{m}\left | \psi_{j}\left ( t \right ) \right |^{4}\leq \left ( \sum_{j=1}^{m}\left | \psi_{j}\left ( t \right ) \right |^{2} \right )^{2}=1.
\end{equation}}
  The IPN is an important physical quantity used to represent the degree of localization of the wave function \cite{localization}. For our multi-flux lattice system, \add{assume the number of unit cells in the lattice is $n$, and the number of paths is $N$.} The specific form of IPN is expanded to
\add{\begin{equation}\label{p^{-1}in our system}
p^{-1}(t)=\sum_{j=1}^{m}\left | \psi_{j}(t) \right |^{4}=\sum_{n}\left | a_{n} \right |^{4}+\sum_{n}\sum_{i=1}^{N}\left | c_{n,i} \right |^{4}.
\end{equation}}
To validate our theoretical framework, we choose the number of paths to be $N=5$ to present the numerical results. In this configuration, the fluxes between the sites under the Aharonov-Bohm constraint must obey
   \begin{equation}\label{phi of N=5}
       \begin{aligned}
    \phi_{n,1}&=0,\ \phi_{n,2}=\phi_{n,3}=\phi,\\
    \phi_{n,4}&=\phi_{n,5}=-\phi.
\end{aligned}
   \end{equation}
Referring to Eq. (\ref{odd condition}), one can obtain $\phi=\arccos(-1/4)$. To systematically investigate the AB caging effect in our one-dimensional photonic lattice ($N=5$), we have performed full numerical simulations and discussions under different transition fluxes. Set the unit of time $t$ in all subsequent numerical simulations is $\pi/J$. We simulated a one-dimensional chain of 9 cells which contains 9 $A_{n}$ and 8 groups of $C_{n,i}$ and totally 49 single sites. \add{Therefore, the IPN at any time can be written as}
\add{\begin{equation}\label{p^{-1}(t)}
p^{-1}(t)=\sum_{n=1}^{9}\left | a_{n} \right |^{4}+\sum_{n=1}^{8}\sum_{i=1}^{5}\left | c_{n,i} \right |^{4}.
\end{equation}}
 Numerical simulations verified that the AB caging effect is insensitive to the specific position of the initial excitation. Consequently, we choose the fifth lattice site as the excitation point and all other sites are in the ground state in this work. \add{At $t=0$, the wavefunction of the lattice only distributes on site $A_{5}$, and the IPN is 1.}

\begin{figure}[tbp]
    \includegraphics[width=1\linewidth]{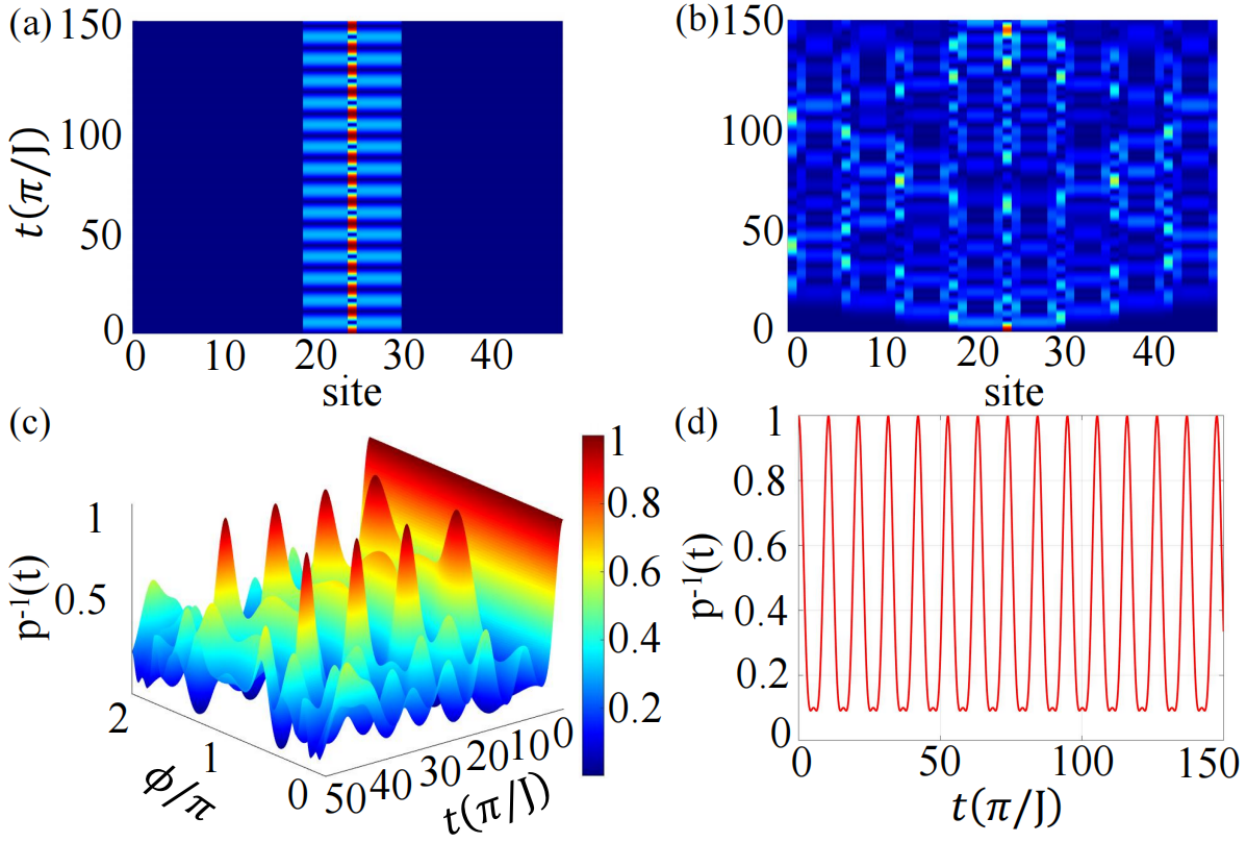}
        \caption{The lattice system we studied consists of 9 unit cells, 49 sites in total, and the initial state of
        $ \left| A_{i} \right\rangle=\left| 0,0...1_{i}...0 \right\rangle$ excited in the lattice. (a) The dynamics process of photon in the lattice reveals the perfect localization phenomenon; (b)The AB caging effect is absent, and the photon presents a transport process in the lattice; (c) The evolution law of IPN with the passage of time; (d) The IPN changes according to the time when the AB caging occurs.}
    \label{f2}
\end{figure}

Fig. \ref{f2} presents the observed key dynamical behaviors and localization characteristics. When the transition flux is set to the critical value $\phi=\arccos(-1/4)$ according to Eq. (\ref{odd condition}), Fig.\;\ref{f2}(a) demonstrates the characteristic AB caging phenomenon. The site-time heat map shows the periodic oscillatory law, and the population of photons at different sites changes periodically. Photons are confined between the sites 20 to 30, indicating that photons show a breathing vibration law near the excitation site. The red position represents the excitation site of photons, and destructive interference prevents energy propagating beyond this area. This completely agrees with our theoretical prediction that the interference effect leads to a complete localization at $\phi=\arccos(-1/4)$. \add{When AB caging occurs in the system, particles exhibit breathing oscillations between lattice site $A_{5}$ and its ten neighboring sites $C_{4,1},C_{4,2}...C_{5,5}$. When the wave function is evenly distributed on the sites $C_{4,1},C_{4,2}...C_{5,5}$, the IPN of the system reads} 
\add{\begin{equation}\label{p^{-1}(t)(cni)}
p^{-1}(t)=\sum_{n=4}^{5}\sum_{i=1}^{5}\left | c_{n,i} \right |^{4}=10\cdot\left | \frac{1}{\sqrt{10}} \right |^{4}=0.1.
\end{equation}}
In contrast, \del{Fig.\;\ref{f2}(b) shows the collapse of the AB caging when $\phi=0$. The dynamics of the photon in the lattice shows all over the entire lattice, illustrating the non-local transport of the photon. The localization of the photons is destroyed when the transition flux does not meet the caging condition, leading the unhindered transport of the photons in the lattice.} \add{Fig.\;\ref{f2}(b) shows the dynamics of photons in the lattice for $\phi=0$. It can be seen that the dynamical evolution of photons spreads across the entire lattice, showing the non-local transport of photons. When the transition flux $\phi$ does not satisfy the AB caging condition, the AB caging effect is absent, so photons can travel freely through the lattice.} Fig.\;\ref{f2}(c) shows the variation of $p^{-1}\left ( t \right )$ according to time $t$ and $\phi$. The sharp red peaks that occur at $\phi=\arccos(-1/4)$ correspond to strong localization, while the flat regions at $\phi\neq\arccos(-1/4)$ indicate that the caging effect does not occur. In Fig.\;\ref{f2}(d), we simulated the evolution of the IPN at $\phi=\arccos(-1/4)$. It further confirms that the cage effect persists over time, causing $p^{-1}\left ( t\right )$ to exhibit period oscillations between \del{0 and 1} \add{0.1 and 1} due to the effect of caging.

The results show that the implementation of $ N = 5$ AB caging in our system requires precise control of the transition flux $\phi = \arccos(-1/4)$. The dynamical process of photons in the lattice and the numerical calculation of the IPN provide directly evidence for tunable localization, which is significant for further research.
\section{AHARONOV-BOHM CAGING BREAKING }
\subsection{Aharonov-Bohm Caging breaking by tunable detuning}
When the transition flux in the lattice meets the AB caging condition, the injection of an on site  potential $\Delta$ at the lattice point usually destroys the degeneracy of system and leads to hybridization between the bands\cite{leakage, disorder}, which leads to the destruction of the localization of photons. However, in the fully flat band case, an inverse Anderson transition can arise, adding \del{disorder} \add{random disorder} breaks the geometric localization of compact states and enables transport in the lattice.

\begin{figure}[htbp]
    \includegraphics[width=1\linewidth]{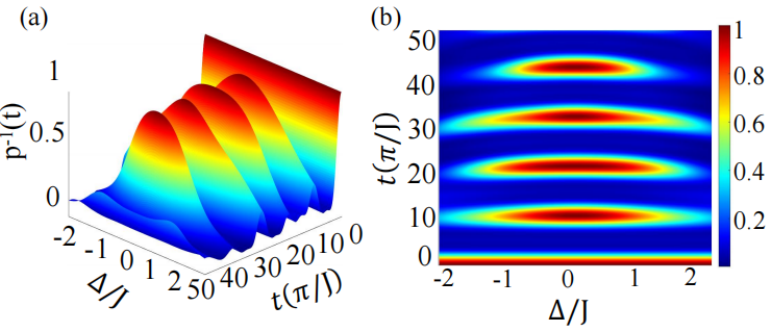}
        \caption{(a) By regulating the strength of \del{disorder} \add{on-site detuning}, the evolution process of the IPN according to time is observed. (b) The stereoscopic picture in (a) is transformed into a plane picture for observation, and the cage effect will be destroyed quickly when the strength of the \del{disorder} \add{on-site detuning} increases, at last the AB caging effect will disappear. }
    \label{f3}
\end{figure}

\add{By contrast, for our system,} in the absence of \del{disorder} \add{on-site detuning}, \del{the system} \add{it} exhibits the typical characteristics of AB caging. When gradually increasing the strength of \del{disorder} \add{on-site detuning} $\Delta$, the AB caging effects is gradually weakening until it disappears. This behavior is a typical characteristic of the inverse Anderson transition, rather than the conventional Anderson localization. In order to quantitatively investigate the influence of the strength of \del{disorder} \add{on-site detuning} on AB caging, we systematically simulated the characteristics of IPN in the case of injecting an energy potential to the lattice points. When an energy potential $\Delta$ is injected into each lattice point $\hat{C}_{n,1}$ and $\hat{C}_{n,5}$, the Hamiltonian of the system can be written as
\begin{equation}     \label{Hsum+Delta}
    \begin{aligned}
   \hat{H}_{\Delta}&=\sum_{n}\sum_{i=1}^{N}\left ( Je^{-i\phi _{i}}\hat{A}_{n}^{\dagger}\hat{C}_{n,i}+J\hat{A}_{n}^{\dagger}\hat{C}_{n-1,i}+\text{H.c.} \right ) \\
   &+\sum_{n}\left ( \Delta\hat{C}^{\dagger}_{n,1}\hat{C}_{n,1}-\Delta\hat{C}^{\dagger}_{n,5}\hat{C}_{n,5}\right ).
\end{aligned}
\end{equation}
The influence of the \del{disorder strength} \add{on-site detuning strength} on the AB caging condition is revealed in the three-dimensional plot shown in Fig.\;\ref{f3}(a), by the variation of $p^{-1}\left ( t\right )$ according to time $t$ and \del{disorder strength} \add{on-site detuning strength} $\Delta/J$. For the regions of very weak \del{disorder} \add{on-site detuning}, the IPN exhibits periodic oscillations, the red peaks represent the strong localization, revealing the process of photon transport in the lattice is hindered, which is the phenomenon of the AB caging. Fig.\;\ref{f3}(b) further confirms this, where the red stripes represent the persistent high IPN. For the weak \del{disorder} \add{on-site detuning} case, the persistent presence of the IPN oscillations in Fig.\;\ref{f3} shows that the effect of AB caging in the weak \del{disorder} \add{on-site detuning} regime is less affected. For the strong \del{disorder} \add{on-site detuning} phase IPN transition is affected rather significant.

\begin{figure}[htbp]
    \includegraphics[width=1\linewidth]{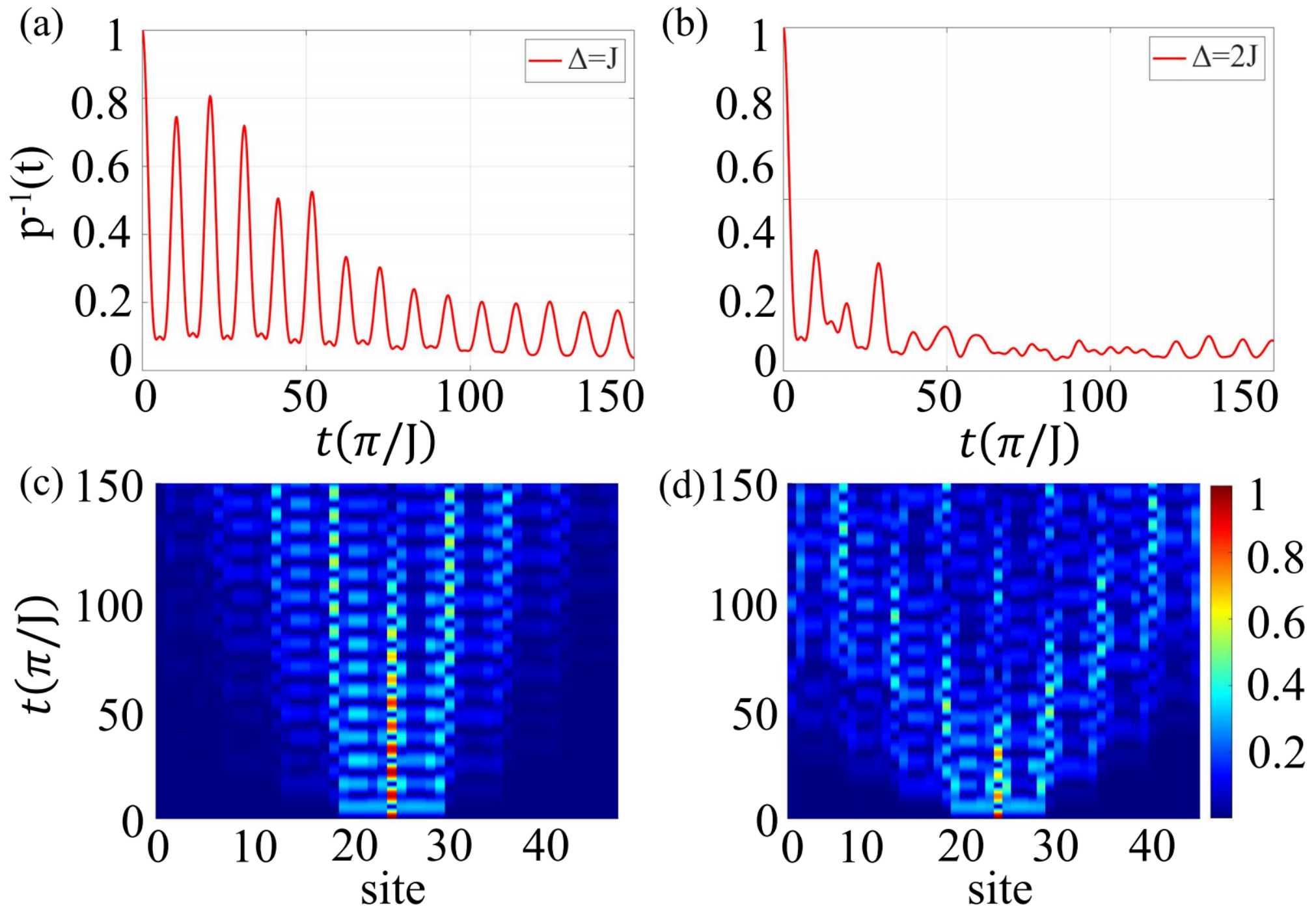}
        \caption{ Time evolution of IPN for the different \del{disorder strength} \add{on-site detuning strength}. The time evolution of IPN for the case: (a) $\Delta=J$ and (b) $\Delta=2J$. The dynamics of photons in the lattice for the case: (c) $\Delta=J$ and (d) $\Delta=2J$.}
    \label{f4}
\end{figure}

 In order to quantitatively study the influence of \del{disorder strength} \add{on-site detuning strength} on AB caging, we have chosen two representative \del{disorder strengths} \add{on-site detuning strengths} $\Delta=J$ and $\Delta=2J$ to carry out numerical simulations of the dynamic process of IPN and photons in the lattice. Both cases show the phenomenon that localization is destroyed after adding \del{disorder} \add{on-site detuning} to the flat band system. For the case of relatively weak \del{disorder} \add{on-site detuning} $\Delta=J$, a relatively long relaxation time is needed for the destruction of AB caging. And for the case of strong \del{disorder} \add{on-site detuning} $\Delta=2J$, the AB caging effect destructs in a very short relaxation time. As shown in Fig.\;\ref{f4}, the evolution of IPN according to time exhibits different kinetic behaviors at different \del{disorder strengths} \add{on-site detuning strengths}. For $\Delta=J$ in Fig.\;\ref{f4}(a), the IPN exhibits periodic oscillations with gradually decreasing amplitude and eventually decays to \del{0} \add{a small non-zero value} from 1 within the observation time. This suggests that the AB caging is destroyed, allowing photon transmission, but the destruction is relatively inefficient. In contrast, for $\Delta=2J$ in Fig.\;\ref{f4}(b), the IPN shows more rapid decay with significantly suppressed oscillation amplitude, which indicates a quick destruction of the caging effect. Moreover, the dynamics process of photons in the lattice also has significant differences under the two \del{disorders} \add{on-site detunings}. The dynamics process of photons in the lattice under weak \del{disorder} \add{on-site detuning} is shown in Fig.\;\ref{f4}(c). It can be found that when the \del{disorder} \add{on-site detuning} $\Delta$ is on the order of magnitude as the coupling $J$, the AB caging is gradually destroyed and the transport process of photons becomes increasingly evident. The dynamics process of photons in the lattice under strong \del{disorder} \add{on-site detuning} is shown Fig.\;\ref{f4}(d). It can be found that when the \del{disorder} \add{on-site detuning} $\Delta$ dominates, the AB caging is rapidly destroyed and the transport range of lattice is relatively large.

\subsection{Ensemble average over multiple independent random disorders}

In this subsection, we introduce two uniform random on-site disorders with different strengths into the ideal flat-band lattice and perform ensemble averages over several independent disorder configurations. We show how disorder strength and the number of samples influence the dynamics of the IPN. These results prove that disorder destroys AB caging, and show the dynamical behavior of AB caging under random disorder. With enough statistical data, these results are more reliable than those in Fig.\;\ref{f4}.

In real physical system, the disorder strength $\Delta$ is usually not a fixed value, but fluctuates randomly within a certain range. To accurately reflect the influence of disorder on AB caging and the localization of the system, we perform an ensemble average over multiple disorder configurations. As described in Eq. (\ref{Hsum+Delta}), we introduce antisymmetric disorder with strengths $\Delta$ and $-\Delta$ on a total of 16 lattice sites, namely $C_{n,1}$ and $C_{n,5}$ in the lattice. For the weak disorder regime, each of these 16 values of $\Delta$ is randomly generated in the range from 0 to $2J$, corresponding to a configuration with a given average disorder strength $\bar{\Delta}=J$. Within a fixed evolution time $T$, each set of disorder yields one dynamical evolution process. To ensure better comparability and reliability of the results, we perform numerical calculations using $N_\mathrm{rep}=500$ independent random disorder configurations respectively, where $N_\mathrm{rep}$ denotes the number of disorder configurations. The final evolution results presented are the ensemble averages over all these configurations. For the strong disorder regime, each of these 16 values of $\Delta$ is randomly generated in the range from 0 to $4J$, corresponding to a configuration with a given average disorder strength $\bar{\Delta}=2J$. We use the same method as above to study the effects of strong disorder on AB caging and the localization of the system. For uniformity, we set $t$(in units of $\pi/J$) to range from 0 to 150 over the total evolution time $T$ of the system.

\begin{figure}[htbp]
    \includegraphics[width=1\linewidth]{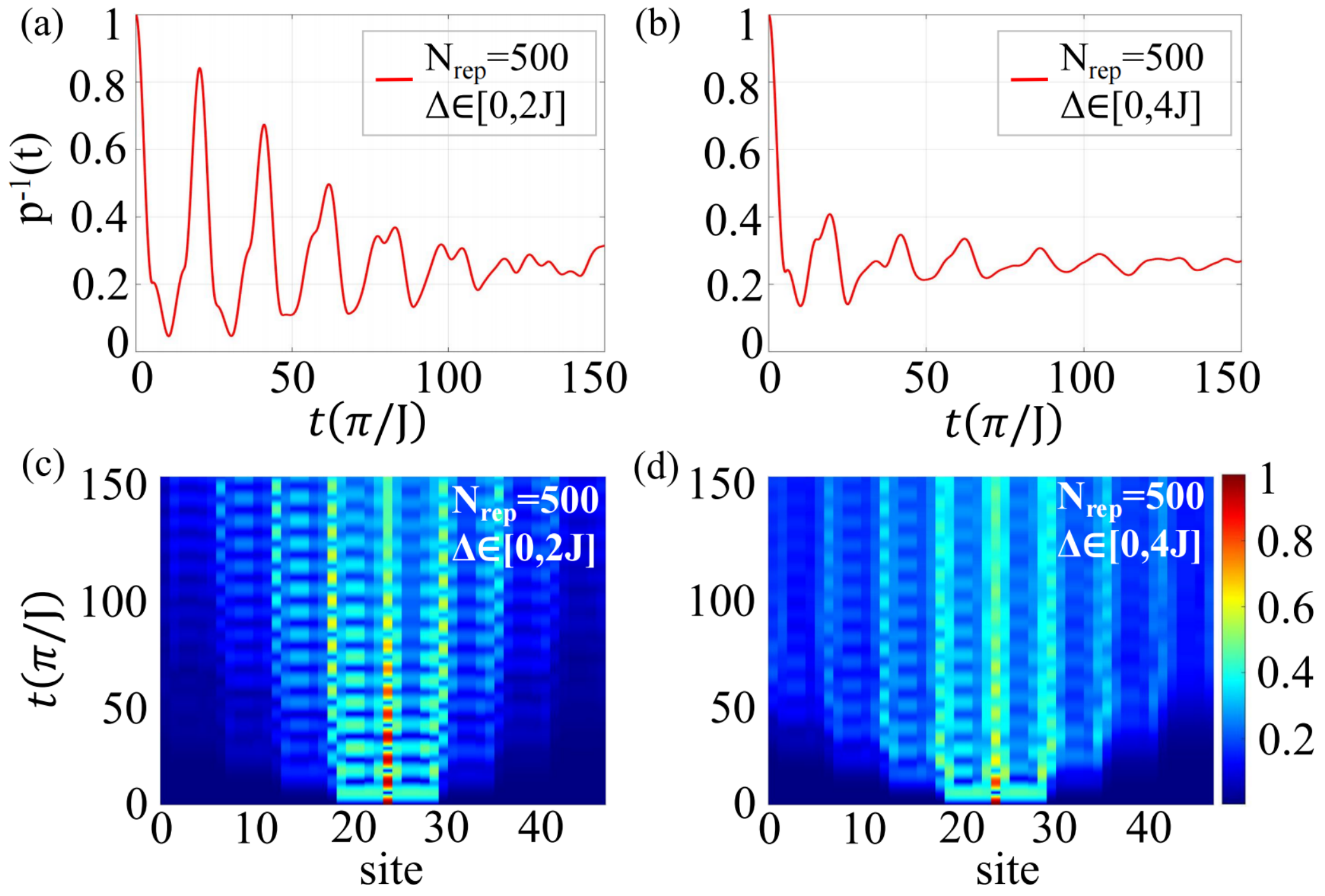}
        \caption{Time evolution of IPN for the different disorder strength and each case is calculated with 500 disorder realizations. The time evolution of IPN for the case: (a) $\Delta$ is randomly distributed between 0 and $2J$. (b) $\Delta$ is randomly distributed between 0 and $4J$. The dynamics of photons in the lattice for the case: (c) $\Delta$ is randomly distributed between 0 and $2J$. (d) $\Delta$ is randomly distributed between 0 and $4J$.}
    \label{f5}
\end{figure}

For the above case, we performed numerical simulations as shown in Fig.\;\ref{f5}, which shows the time evolution of the IPN after introducing disorder into the ideal flat-band lattice and the dynamical process of particles in the ideal flat-band lattice after introducing disorder. By comparing Fig.\;\ref{f5}(a) and (b), it can be seen that stronger disorder leads to faster destruction of the system localization. By comparing Fig.\;\ref{f5}(c) and (d), it can be seen that stronger disorder increases the transport rate of particles in the lattice, leading to faster destruction of AB caging. In general, these results in Fig. \ref{f5} are more convincing than those in Fig. \ref{f4}. It shows that the stronger disorder accelerates the destruction of the system's localization and AB caging effect, and also speeds up the particle transport in the lattice.

\subsection{The fluctuations of inverse participation number and the localization characteristics}
In the previous discussion, we characterize the localization degree of the system mainly using the IPN. Our results show that the IPN exhibits periodic oscillations between \del{0 and 1} \add{0.1 and 1} when AB caging occurs. However, when considering the combined effects of two factors on the stability of AB caging and the localization characteristics, the IPN exhibits significant time dependence, making it difficult to fully characterize the system behavior only using the instant IPN. So we introduce the fluctuation of the IPN as a localization criterion from a statistical perspective, and  analyze the combined influences of the two factors on the stable-state behavior of AB caging more systematically.

The average value and the mean square value of the IPN over the evolution time $T$ can be written as
\begin{equation}
    \begin{aligned}\label{平均值}
    \overline{IPN}&=\frac{1}{T}\int_{0}^{T}p^{-1}\left ( t \right )\mathrm{d}t\;,\\
  \overline{\left ( IPN \right )^{2}}&=\frac{1}{T}\int_{0}^{T}[ p^{-1}\left ( t \right ) ]^{2}\mathrm{d}t\;.
\end{aligned}
\end{equation}
Thus, the fluctuation of the IPN can be written as
\begin{equation}\label{涨落}
    \sigma=\sqrt{\overline{\left ( IPN \right )^{2}}-\left ( \overline{IPN} \right )^{2}}\;.
\end{equation}

\begin{figure}[htbp]
    \includegraphics[width=1\linewidth]{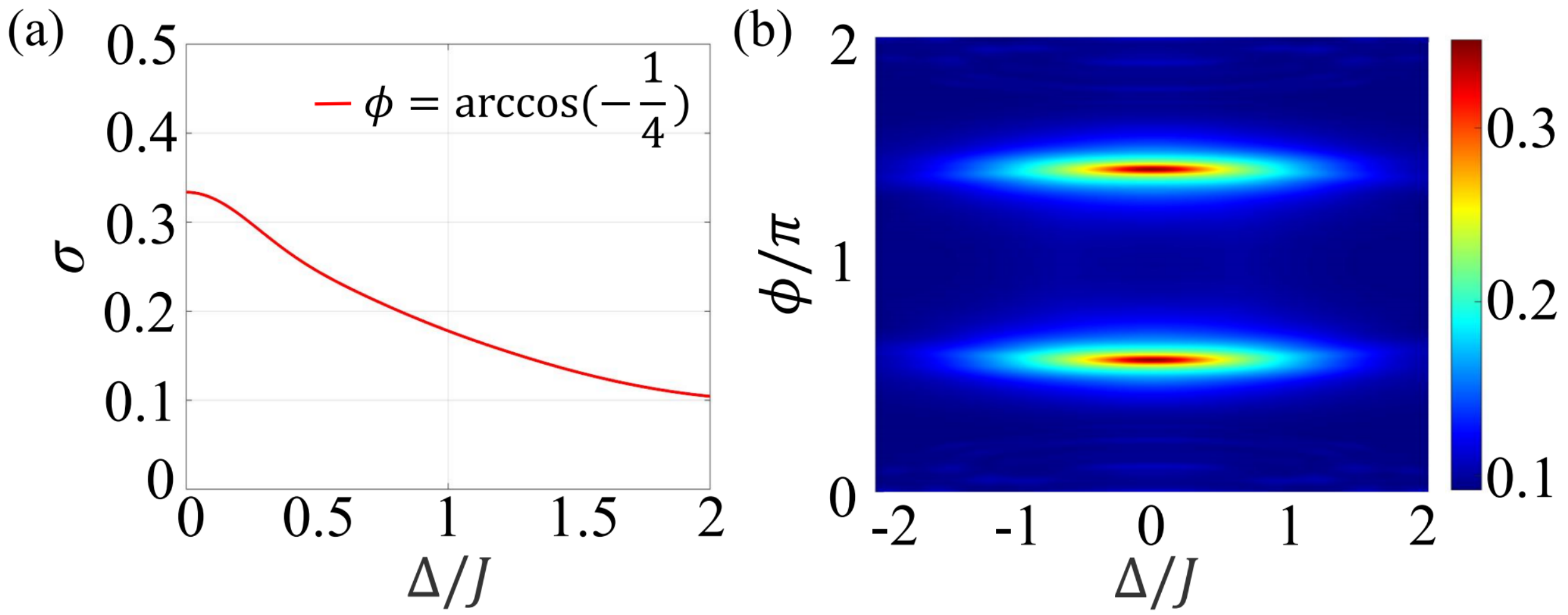}
        \caption{(a) The variation of the IPN fluctuation $\sigma$ with \del{disorder} \add{on-site detuning} $\Delta$ under the flat-band lattice condition. (b) By controlling both the flux $\phi$ and \del{disorder} \add{on-site detuning} $\Delta$, we observe the dependence of the IPN fluctuation $\sigma$ on these two parameters. The phase diagram in the ($\Delta$, $\phi$) plane illustrates how the flux $\phi$ and \del{disorder} \add{on-site detuning} $\Delta$ influence the stability of AB caging.}
    \label{f6}
\end{figure}

The fluctuation can reflect the deviation of the AB caging effect near the steady state. As a specific example, we analyze the combined effect of flux $\phi$ and \del{disorder} \add{on-site detuning} $\Delta$ on the AB caging effect. As shown in Fig.\;\ref{f6}, we investigate the variation of the IPN fluctuation $\sigma$ with \del{disorder} \add{on-site detuning} $\Delta$ under the flat-band lattice condition. In addition, we numerically obtain the dependence of the IPN fluctuation $\sigma$ on $\phi$ and $\Delta$, and accordingly construct the phase diagram in the ($\Delta$, $\phi$) parameter plane. For consistency, we performed numerical calculations under the same conditions as above. Regarding the choice of evolution time $T$, we set $t$(in units of $\pi/J$) to range from 0 to 150, and adopted 300 points over the entire evolution time $T$ to calculate the IPN fluctuation $\sigma$.

Fig.\;\ref{f6}(a) presents the variation of the IPN fluctuation $\sigma$ with \del{disorder} \add{on-site detuning} $\Delta$ at a fixed flux $\phi = \arccos(-1/4)$. When $\Delta=0$, the system exhibits the AB caging effect, corresponding to the maximum value of the IPN fluctuation. As the \del{disorder} \add{on-site detuning} strength gradually increases, the localization induced by AB caging is progressively destroyed, leading to a gradual reduction in both the IPN and its fluctuation. Since the IPN fluctuation $\sigma$ is calculated over the entire evolution time $T$, this picture also indirectly reflects how the time scale of IPN decay changes with \del{disorder} \add{on-site detuning} strength, and can be viewed as a scaling law describing the degree of localization destruction as a function of \del{disorder} \add{on-site detuning} strength.

Fig.\;\ref{f6}(b) shows the combined effects of flux $\phi$ and \del{disorder} \add{on-site detuning} $\Delta$ on AB caging using a phase diagram in the ($\Delta$, $\phi$) plane. The phase diagram reflects the AB caging is highly sensitive to the variations in both flux $\phi$ and \del{disorder} \add{on-site detuning} $\Delta$. The value $\phi=\arccos(-1/4)$ is where AB caging occurs in this system. Along the two vertical lines at $\phi=\arccos(-1/4)$, it can be observed that the IPN fluctuation $\sigma$ first increases and then decreases as the \del{disorder} \add{on-site detuning} $\Delta$ varies from $-2J$ to $2J$. The fluctuation reaches its maximum at $\Delta=0$, which is the ideal AB caging state, where the localization is strongest. When $\phi\neq\arccos(-1/4)$, the IPN fluctuation stays near its minimum value, because the AB caging is absent.

\subsection{Dissipative effects on Multi-Flux Aharonov-Bohm caging effect}
In this subsection, we discuss the effect of system dissipation on the stability of the AB caging effect. Since all sites in our system are excited at single lattice sites, we introduce virtual state $\left| v \right\rangle$ to form the zero-excitation subspace. The occupation numbers on all lattice sites decay into this newly defined zero-excitation subspace at the same rate $\gamma/J$.

The k-th dissipation operator of the system can be written as $\hat{\mathcal{L}}_{k}=\sqrt{\gamma}\left| v \right\rangle\left \langle k \right |$, it describes the decay process of the k-th lattice site (where k = 1, 2, ..., 49) in the lattice to the virtual state. Setting $\hbar=1$ and considering the effect of dissipation, the evolution of the system density operator $\hat{\rho}$ is described by the Lindblad master equation,
\begin{equation}\label{master equation}
\frac{\mathrm{d} \hat{\rho}}{\mathrm{d} t}=-i\left [ \hat{H},\hat{\rho} \right ]+\sum_{k}\left ( \hat{\mathcal{L}}_{k}\hat{\rho}\hat{\mathcal{L}}^{\dagger}_{k}-\frac{1}{2} \left \lbrace \hat{\mathcal{L}}^{\dagger}_{k}\hat{\mathcal{L}}_{k},\hat{\rho} \right \rbrace\right ).
\end{equation}

\begin{figure}[htbp]
    \includegraphics[width=1\linewidth]{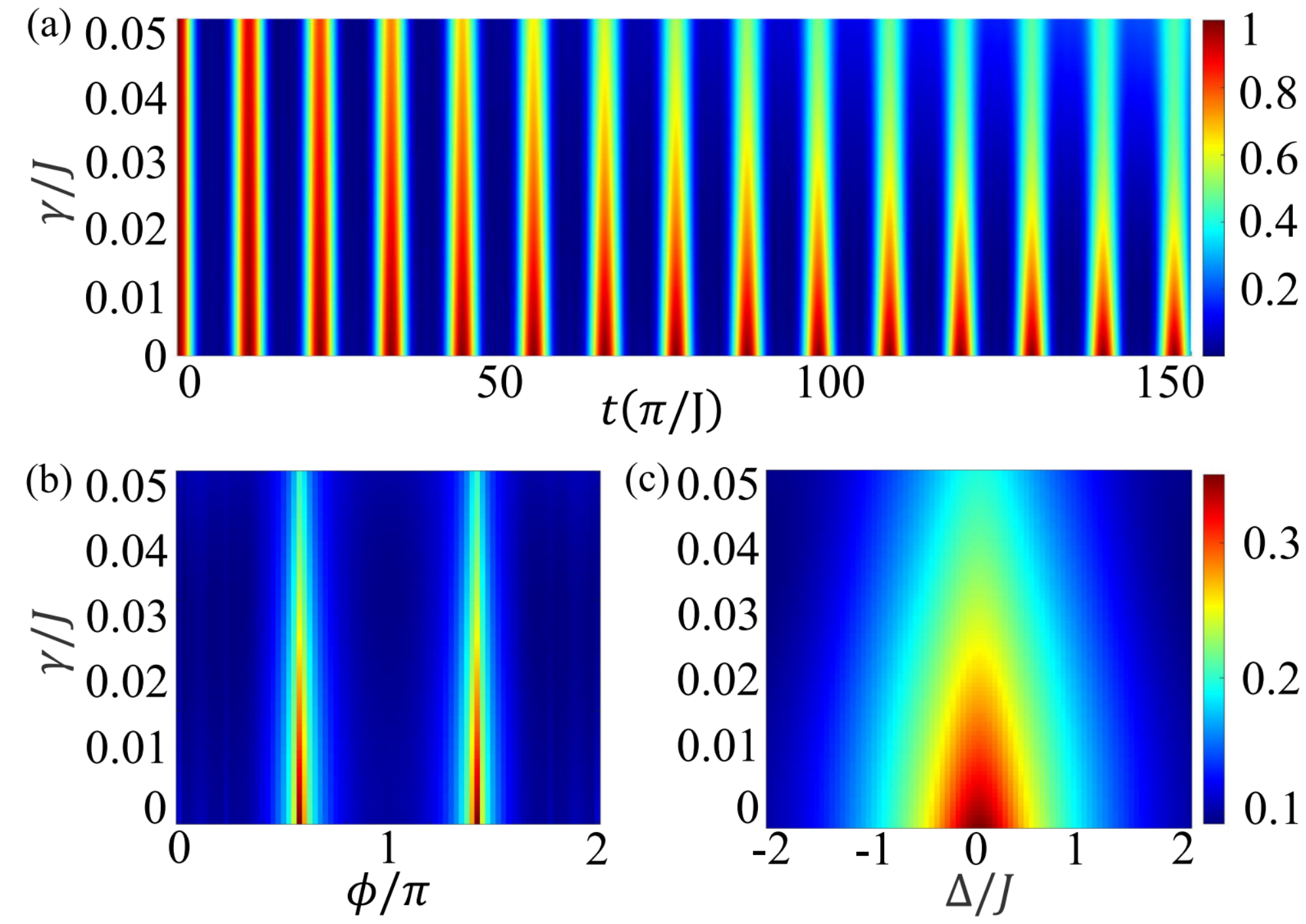}
        \caption{(a) By regulating the strength of dissipation, the evolution process of the IPN according to time is observed. (b) The relation of the IPN fluctuation $\sigma$ as a function of dissipation $\gamma$ and flux $\phi$. (c) The relation of the IPN fluctuation $\sigma$ as a function of dissipation $\gamma$ and \del{disorder} \add{on-site detuning} $\Delta$.}
    \label{f7}
\end{figure}

To show the general influence of dissipation on AB caging and the robustness of AB caging against dissipation and other physical factors, we performed numerical simulations as shown in Fig.\;\ref{f7}.

Fig.\;\ref{f7}(a) shows the influence of dissipation rate on AB caging, we control the dissipation $\gamma$ within the range of 0 to $0.05J$. Under weak or even zero dissipation, the IPN exhibits periodic oscillations over time, where the red peaks represent strongly localized regions, indicating that the photon transport process in the lattice is prevented. This means that the destruction of the localization of the system is quite slight. In the regime of moderate or strong dissipation, AB caging effect and the degree of system localization is observed to be significantly destroyed according to the time.

Fig.\;\ref{f7}(b) shows the combined effects of flux $\phi$ and dissipation $\gamma$ on AB caging using a phase diagram in the plane ($\gamma$, $\phi$). The phase diagram reflects the robustness of AB caging against variations in $\phi$ and $\gamma$, and it can be seen that AB caging is highly sensitive to the changes in both the flux $\phi$ and the dissipation $\gamma$. When the flux $\phi$ satisfies the condition for AB caging, the fluctuation of the IPN decreases as the dissipation rate increases from 0 to $0.05J$, during which the localization of the system is gradually destroyed. When the flux $\phi$ does not satisfy the condition for AB caging, AB caging is absent and the degree of localization of the system remains weak throughout. With the gradual increases of dissipation $\gamma$, the fluctuation of IPN does not change significantly. Fig.\;\ref{f7}(c) shows the effect of introducing both \del{disorder} \add{on-site detuning} $\Delta$ and dissipation $\gamma$ on AB caging with a phase diagram in the ($\Delta$, $\gamma$) plane. The phase diagram reflects the robustness of AB caging against variations in $\Delta$ and $\gamma$, and it can be seen that AB caging is also highly sensitive to the changes in both \del{disorder} \add{on-site detuning} $\Delta$ and dissipation $\gamma$. The maximum fluctuation of the system occurs exactly at the point where both $\Delta$ and $\gamma$ are zero, which represents ideal AB caging under $\phi=\arccos(-1/4)$. When the dissipation rate increases, some particles decay to the virtual state more rapidly, and the localization of the virtual state is enhanced, leading to a slight recovery of the IPN and its fluctuation. It can be further concluded that as the dissipation rate increases, a large number of particles in the lattice will decay into the virtual state more rapidly over time. After a long enough evolution time, all particles in the system will decay into the virtual state. Then, the localization of the virtual state becomes strongest, while the original AB caging of the system is destroyed.

\subsection{The stability of the Multi-Flux Aharonov-Bohm caging effect under the non-Hermitian hopping terms}
 In this subsection, we discuss the effect of introducing the non-Hermitian hopping terms into Eq. (\ref{Hsum}) on AB caging\cite{Non-Hermitian1,Non-Hermitian2,Non-Hermitian3}. We consider non-Hermitian hopping terms between the nearest-neighbor sites. The total Hamiltonian of this system can be written as
\begin{equation}     \label{Hsum+i}
   \begin{aligned}
   \hat{H}_{\alpha}&=\sum_{n}\sum_{i=1}^{N}\left ( Je^{-i\phi _{i}}\hat{A}_{n}^{\dagger}\hat{C}_{n,i}+J\hat{A}_{n}^{\dagger}\hat{C}_{n-1,i}+\text{H.c.} \right )\\
 &-i\Gamma \sum_{n}\left ( \hat{A}_{n}^{\dagger}\hat{C}_{n,i}+ \hat{A}_{n}^{\dagger}\hat{C}_{n-1,i}+\text{H.c.}\right ).
 \end{aligned}
\end{equation}
Here, $\Gamma$ is the non-Hermitian hopping strength between the nearest-neighbor sites, and the total Hamiltonian $\hat{H}_{\alpha}$ is non-Hermitian. Now we performed numerical simulations and discussion with exactly the same method as before.
Notably, we only consider the effect of the non-Hermitian hopping terms in this subsection and do not include dissipation.
\begin{figure}[htbp]
    \includegraphics[width=1\linewidth]{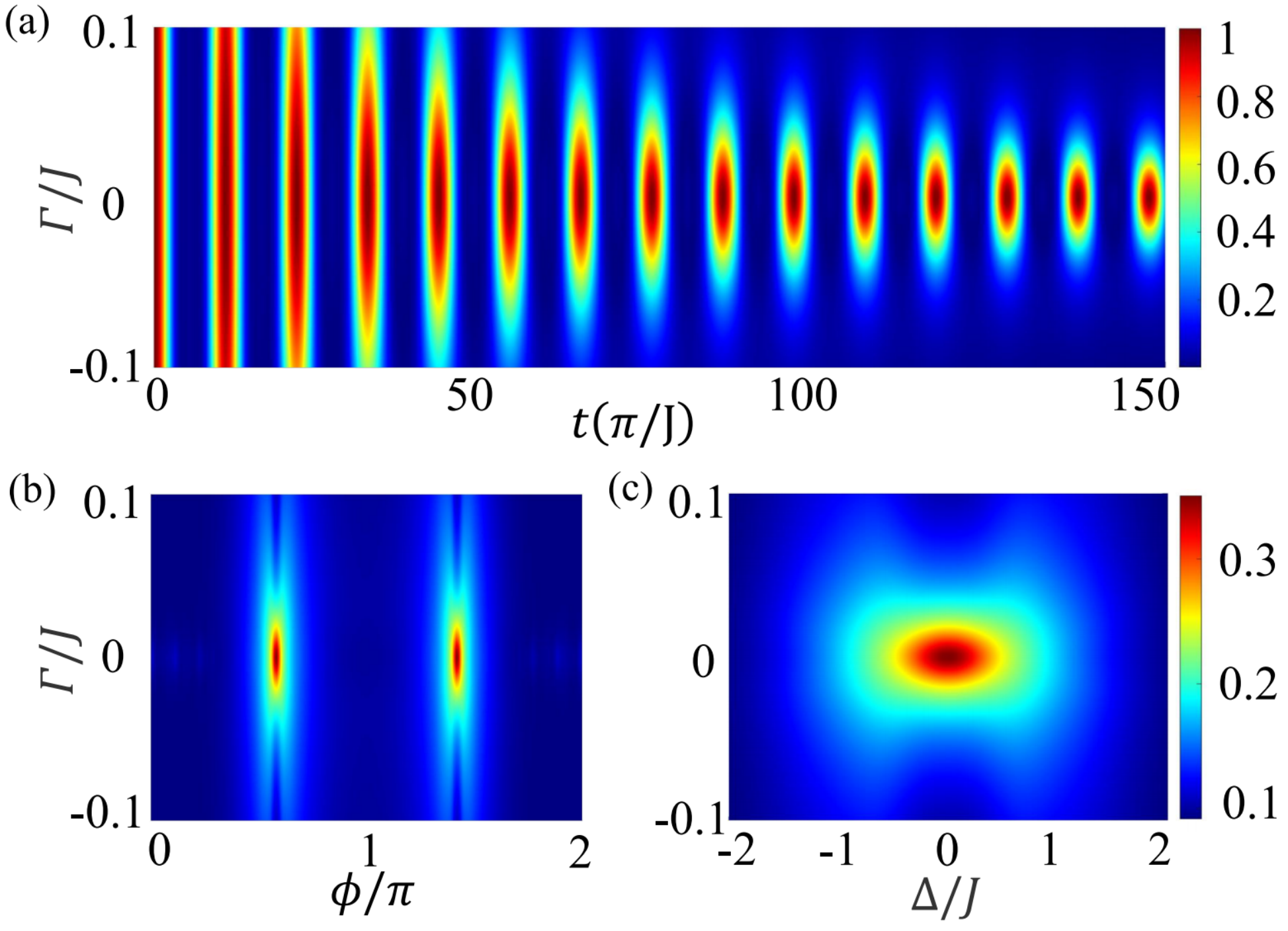}
        \caption{(a) By regulating the strength of non-Hermitian hopping terms, the evolution process of the IPN according to time is observed. (b) The relation of the IPN fluctuation $\sigma$ as a function of non-Hermitian hopping $\Gamma$ and flux $\phi$. (c) The relation of the IPN fluctuation $\sigma$ as a function of non-Hermitian hopping $\Gamma$ and \del{disorder} \add{on-site detuning} $\Delta$.}
    \label{f8}
\end{figure}

To show the general influence of non-Hermitian hopping terms on AB caging and the robustness of AB caging against non-Hermitian hopping and other physical factors, we performed numerical simulations as shown in Fig.\;\ref{f8}.

Fig.\ref{f8}(a) shows the effect of non-Hermitian hopping strength on AB caging, where $\Gamma$ varies from $-0.1J$ to $0.1J$. When the non-Hermitian hopping strength is weak or even zero, the IPN oscillates periodically with time. The red peaks represent strongly localized regions, indicating that the transport of photons in the lattice is prevented. This means that the localization of the system is slightly destroyed or remains intact. In regions with strong non-Hermitian hopping strength, the AB caging and the localization of the system are significantly destroyed as time increases.

Fig.\;\ref{f8}(b) presents a phase diagram in the ($\Gamma$, $\phi$) plane, showing the combined effect of the flux $\phi$ and non-Hermitian hopping $\Gamma$ on AB caging. This phase diagram also reflects the robustness of AB caging against variations in $\phi$ and $\Gamma$, and it can be observed that AB caging is highly sensitive to the variations of these two parameters. When the flux $\phi$ satisfies the condition for AB caging, we can see along the two vertical lines corresponding to $\phi=\arccos(-1/4)$ that the fluctuation $\sigma$ of the IPN first increases and then decreases as the non-Hermitian hopping strength increases from $-0.1J$ to $0.1J$. At $\Gamma=0$, the fluctuation reaches its maximum, relating to the ideal AB caging effect and the strongest localization of the system. In regions where $\phi\neq\arccos(-1/4)$, the fluctuation of the IPN remains near a minimum value because AB caging is absent. The localization of the system remains weak, and the fluctuation of the IPN is always small regardless of how the non-Hermitian hopping $\Gamma$ changes.

Fig.\;\ref{f8}(c) presents a phase diagram in the ($\Delta$, $\Gamma$) plane, showing the combined effect of \del{disorder} \add{on-site detuning} $\Delta$ and non-Hermitian hopping $\Gamma$ on AB caging in a flat-band lattice. This phase diagram reflects the robustness of AB caging against the variations in $\Delta$ and $\Gamma$, and it can be seen that AB caging is also highly sensitive to the changes in these two parameters. The maximum fluctuation of the system occurs exactly at the point where both \del{disorder} \add{on-site detuning} $\Delta$ and non-Hermitian hopping $\Gamma$ are zero, representing the ideal AB caging state. When $\Gamma=0$ and $\Delta$ varies from $-2J$ to $2J$, the fluctuation of the IPN first increases and then decreases, indicating that the system localization is enhanced first and then weakened. When $\Delta=0$ and $\Gamma$ varies from $-0.1J$ to $0.1J$, the fluctuation of the IPN shows the same behavior. It can be observed that when either $\Delta$ or $\Gamma$ increases, or both increase together, the destruction of AB caging is cumulative. These two factors act cooperatively to break the system localization.

\section{PHYSICAL REALIZATION IN A SUPERCONDUCTING CIRCUITS SYSTEM}
\begin{figure}[htbp]
    \includegraphics[width=1\linewidth]{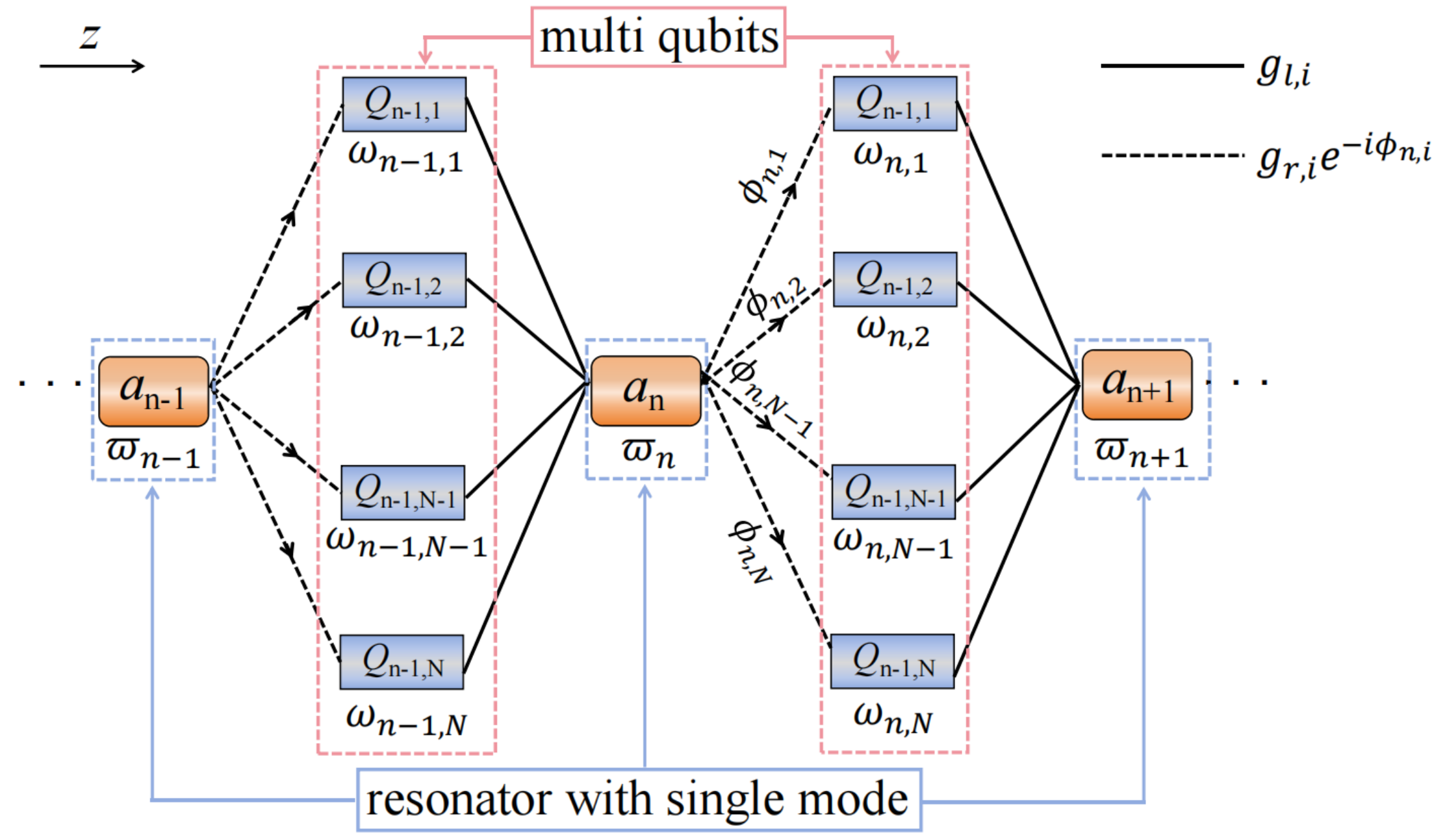}
        \caption{The interaction between multiple qubits and a resonator is shown in this figure, where the solid line represents the coupling strength $g_{l,i}$. The dashed line indicates that there is a controllable phase $\phi_{n,i}$ in the coupling between the lattice, and there are $N$ paths ($N$ qubits) between the adjacent resonators $a_{n}$ and $a_{n+1}$. The transmission process of photons in the lattice is along the $z$ direction.}
    \label{f9}
\end{figure}

In this section, we present the realization of our scheme on superconducting quantum circuits. On the superconducting quantum circuits, the coupling between single-mode electromagnetic field in the resonator and qubits is very common in the circuit QED architecture. The multi-flux lattice model in  Fig.\;\ref{f1}  can be described by the coupling between a single-mode cavity and multiple qubits, as shown in Fig.\;\ref{f9}. 

Under the rotating-wave approximation, the Hamiltonian describing the interaction between the n-th cavity and N qubits with distinct frequencies on its left and right sides follows the Jaynes-Cummings (JC) model. Setting $\hbar=1$, the total Hamiltonian of the system is expressed as
\begin{equation}
    \begin{aligned}\label{QED总哈密顿量}
  \hat{H}_{\mathrm{Q}}&=\hat{H}_{Q_{0}}+\hat{V}_{Q}\;,\\
   \hat{H}_{Q_{0}}&=\sum_{n}\sum_{i=1}^{N}\frac{1}{2}\omega_{n,i}\hat{\sigma}_{z}^{\left ( n,i \right )}+\sum_{n}\varpi_{n}\hat{a}^{\dagger}_{n}\hat{a}_{n}\;,\\
    \hat{V}_{Q}&=\sum_{n}\sum_{i=1}^{N}\left ( g_{r,i}\hat{a}_{n}\hat{\sigma}_{+}^{\left ( n,i \right )}e^{-i\phi_{n,i}}+ g_{l,i}\hat{a}_{n-1}\hat{\sigma}_{+}^{\left ( n,i \right )}+\mathrm{H.c.} \right ).
\end{aligned}
\end{equation}

$ \hat{H}_{Q_{0}}$ is the free Hamiltonian of this system. The first term represents the free Hamiltonian of the qubits, where $\omega_{n,i}$ represents the transition frequency from the ground state $\left| g \right\rangle_{n,i}$ to the excited state $\left| e \right\rangle_{n,i}$ for the i-th qubit coupled to the n-th cavity. The second term corresponds to the free Hamiltonian of the cavities, with $\varpi_{n}$ being the resonant frequency of the operating mode of the n-th cavity. $\hat{V}_{Q}$ describes the interaction between qubits and cavities. Here, $g_{r,i}$ is the coupling strength between the i-th qubit and the n-th cavity, and $g_{l,i}$ denotes the coupling strength between the i-th qubit and the (n-1)-th cavity. Meanwhile, a controllable phase $\phi_{n,i}$ can be introduced using Floquet parametric modulation \cite{superconducting_circuits}. 

The Hamiltonian of the system in the interaction picture is written as
\begin{equation}
    \begin{aligned}\label{QED相互作用绘景}
    \hat{V}_{Q_{I}}\left ( t \right ) & = e^{i \hat{H}_{Q_{0}}t}\hat{V}_{Q}e^{-i \hat{H}_{Q_{0}}t} \\
         &= \sum_{n}\sum_{i=1}^{N}\left ( g_{r,i}\hat{a}_{n}\hat{\sigma}_{+}^{\left ( n,i \right )}e^{-i\phi_{n,i}}e^{i\Delta_{n} t}+\mathrm{H.c.}  \right )\\
         &+\sum_{n}\sum_{i=1}^{N}\left ( g_{l,i}\hat{a}_{n-1}\hat{\sigma}_{+}^{\left ( n,i \right )}e^{-i\Delta_{n-1} t}+\mathrm{H.c.} \right ).
\end{aligned}
\end{equation}

Here, $\Delta_{n}=\omega_{n,i}-\varpi_{n}$ is the detuning between the i-th qubit and the n-th cavity. For simplicity, we assume that all qubits share an identical coupling strength g with the cavities. Under the resonant condition $\omega_{n,i}=\varpi_{n}$, Eq. (\ref{QED相互作用绘景}) becomes
\begin{equation} \label{QED相互作用绘景大失谐}
     {\hat{V}}'_{Q_{I}}=\sum_{n}\sum_{i=1}^{N}g\left ( \hat{a}_{n}\hat{\sigma}_{+}^{\left ( n,i \right )}e^{-i\phi_{n,i}}+ \hat{a}_{n-1}\hat{\sigma}_{+}^{\left ( n,i \right )}+\mathrm{H.c.} \right ).
\end{equation}

Eq. (\ref{QED相互作用绘景大失谐}) shares a similar formal structure as Eq. (\ref{Hsum}). Here, $\hat{a}_{n}$ corresponds to the annihilation operator $\hat{A}_{n}$ at lattice site $A_{n}$, and $\hat{\sigma}_{+}^{\left ( n,i \right )}$ corresponds to the creation operator $\hat{C}^{\dagger}_{n,i}$ at lattice site $C_{n,i}$. By choosing appropriate values of qubits $N$ and tuning $\phi_{n,i}$, multi-flux AB caging can be observed in the QED system. \add{Importantly, the relation between the qubit raising operator $\hat{\sigma}_{+}^{\left ( n,i \right )}$ and the bosonic creation operator $\hat{C}^{\dagger}_{n,i}$ holds only in the single-excitation subspace, and this is enough to describe the single-particle AB caging effect.}
\section{CONCLUSIONS}
In summary, we propose a scheme to derive universal conditions for AB caging with
multi-flux. The numerical simulations demonstrate the AB caging effect in our synthetic system under appropriate parameters, showing oscillations over time. We also discuss the robustness of the AB caging effect under \del{disorder} \add{on-site detuning}. It is shown that the intensity of the \del{disorder} \add{on-site detuning} determines the degree and the speed of the destruction of the AB caging.
Furthermore, we perform ensemble average over multiple independent random disorders, and explore the influence of decoherence effects and non-Hermitian
transitions on AB caging. It is shown that the multiple independent random disorders strength, decoherence rate and non-Hermitian transition strength also affect the degree and the speed of the destruction of AB caging.
The multi-flux proposal introduces more parameters, making our approach more tunable and shedding light on the realization of the AB caging effect in suitable physical systems. That may exploit potential applications for related physical device design in the future.
\acknowledgements
We would like to thank M.-J. Liang for helpful discussions. This work was supported by the National Natural Science Foundation of China (Grant No.~92576110), the Guangdong Provincial Quantum Science Strategic Initiative (Grants No.~GDZX2203001 and No.~GDZX2403002), the Guangdong Provincial Key Laboratory (Grant No.~2020B1212060066),  and the Guangdong Project (Grant No.~2024TQ08A960).

\end{document}